\documentstyle[epsf]{mn}

\newif\ifAMStwofonts
\AMStwofontstrue

\newcommand{\be}{\begin{equation}}
\newcommand{\ee}{\end{equation}}
\newcommand{\msol}{\mbox{$\mathrm{M}_{\sun}$}}

\newcommand{\thoc}{$\theta^{1}$C Ori}

\ifoldfss
  \ifCUPmtlplainloaded \else
    \NewTextAlphabet{textbfit} {cmbxti10} {}
    \NewTextAlphabet{textbfss} {cmssbx10} {}
    \NewMathAlphabet{mathbfit} {cmbxti10} {} % for math mode
    \NewMathAlphabet{mathbfss} {cmssbx10} {} %  "   "    "
  \fi
  \ifAMStwofonts
    \ifCUPmtlplainloaded \else
      \NewSymbolFont{upmath} {eurm10}
      \NewSymbolFont{AMSa} {msam10}
      \NewMathSymbol{\upi}     {0}{upmath}{19}
      \NewMathSymbol{\umu}     {0}{upmath}{16}
      \NewMathSymbol{\upartial}{0}{upmath}{40}
      \NewMathSymbol{\leqslant}{3}{AMSa}{36}
      \NewMathSymbol{\geqslant}{3}{AMSa}{3E}

    \fi
  \fi
\fi % End of OFSS

\ifnfssone
  \newmathalphabet{\mathit}
  \addtoversion{normal}{\mathit}{cmr}{m}{it}
  \addtoversion{bold}{\mathit}{cmr}{bx}{it}
  \newmathalphabet{\mathbfit} % math mode version of \textbfit{..}
  \addtoversion{normal}{\mathbfit}{cmr}{bx}{it}
  \addtoversion{bold}{\mathbfit}{cmr}{bx}{it}
  \newmathalphabet{\mathbfss} % math mode version of \textbfss{..}
  \addtoversion{normal}{\mathbfss}{cmss}{bx}{n}
  \addtoversion{bold}{\mathbfss}{cmss}{bx}{n}
  \ifAMStwofonts
    \ifCUPmtlplainloaded \else
      \UseAMStwoboldmath
      \makeatletter
      \new@mathgroup\upmath@group
      \define@mathgroup\mv@normal\upmath@group{eur}{m}{n}
      \define@mathgroup\mv@bold\upmath@group{eur}{b}{n}
      \edef\UPM{\hexnumber\upmath@group}
      \new@mathgroup\amsa@group
      \define@mathgroup\mv@normal\amsa@group{msa}{m}{n}
      \define@mathgroup\mv@bold\amsa@group{msa}{m}{n}
      \edef\AMSa{\hexnumber\amsa@group}
      \makeatother
      \mathchardef\upi="0\UPM19
      \mathchardef\umu="0\UPM16
      \mathchardef\upartial="0\UPM40
      \mathchardef\leqslant="3\AMSa36
      \mathchardef\geqslant="3\AMSa3E
    \fi
  \fi
\fi % End of NFSS release 1

\ifnfsstwo
  \DeclareMathAlphabet{\mathbfit}{OT1}{cmr}{bx}{it}
  \SetMathAlphabet\mathbfit{bold}{OT1}{cmr}{bx}{it}
  \DeclareMathAlphabet{\mathbfss}{OT1}{cmss}{bx}{n}
  \SetMathAlphabet\mathbfss{bold}{OT1}{cmss}{bx}{n}
  \ifAMStwofonts
    \ifCUPmtlplainloaded \else
      \DeclareSymbolFont{UPM}{U}{eur}{m}{n}
      \SetSymbolFont{UPM}{bold}{U}{eur}{b}{n}
      \DeclareSymbolFont{AMSa}{U}{msa}{m}{n}
      \DeclareMathSymbol{\upi}{0}{UPM}{"19}
      \DeclareMathSymbol{\umu}{0}{UPM}{"16}
      \DeclareMathSymbol{\upartial}{0}{UPM}{"40}
      \DeclareMathSymbol{\leqslant}{3}{AMSa}{"36}
      \DeclareMathSymbol{\geqslant}{3}{AMSa}{"3E}
    \fi
  \fi
\fi % End of NFSS release 2

\ifCUPmtlplainloaded \else
  \ifAMStwofonts \else % If no AMS fonts
    \def\upi{\pi}
    \def\umu{\mu}
    \def\upartial{\partial}
  \fi
\fi

\begin{document}

\title[Disk destruction in the Orion Nebula Cluster]{Destruction of
protoplanetary disks in the Orion Nebula Cluster}

\author[A. Scally and C. Clarke]{Aylwyn Scally and Cathie Clarke\\
Institute of Astronomy, Madingley Road, Cambridge CB3 0HA, England}

\maketitle

\begin{abstract}
We use numerical $N$-body simulations of the Orion Nebula Cluster (ONC)
to investigate the destruction of protoplanetary disks by close stellar
encounters and UV radiation from massive stars.  The simulations model
a cluster of 4000 stars, and we consider separately cases in which the
disks have fixed radii of 100 AU and 10 AU. In the former case,
depending on a star's position and orbit in the cluster over $10^7$
years, UV photoevaporation removes at least 0.01 \msol\ from its disk,
and can remove up to 1 \msol. We find no dynamical models of the ONC
consistent with the suggestion of St\"{o}rzer and Hollenbach
\shortcite{StoHol99} that the observed distribution and abundance of
proplyds could be explained by a population of stars on radial orbits
which spend relatively little time near \thoc\ (the most massive star
in the ONC).  Instead the observations require either massive disks
(e.g. a typical initial disk mass of 0.4 \msol) or a very recent birth
for \thoc. When we consider the photoevaporation of the inner 10 AU of
disks in the ONC, we find that planet formation would be hardly
affected. Outside that region, planets would be prevented from forming
in about half the systems, unless either the initial disk masses were
very high (e.g.\ 0.4 \msol) or they formed quickly (in less than $\sim
2$ Myr) and \thoc\ has only very recently appeared.

We also present statistics on the distribution of minimum stellar
encounter separations. This peaks at $1000\,\mbox{AU}$, with only about
4 per cent of stars having had an encounter closer than 100 AU at the
cluster's present age, and less than 10 per cent after $10^7$ years.
We conclude that stellar encounters are unlikely to play a significant
role in destroying protoplanetary disks. In the absence of any
disruption mechanism other than those considered here, we would thus
predict planetary systems like our own to be common amongst stars
forming in ONC-like environments.

Also, although almost all stars will have experienced an encounter at
the radius of the Oort cloud in our own system, this only places a firm
constraint on the possible birthplace of the Sun if the Oort cloud
formed in situ, rather than through the secular ejection of matter from
the planetary zone.
\end{abstract}


\section{Introduction}

The formation of planets from the gas and dust around a young star is
thought to take several million years (e.g.\ Lissauer
\shortcite{Lis93}, Pollack et al. \shortcite{Pol+96}). During this
time, the star is unlikely to move far from the region where it itself
was formed, and its early evolution will continue to be influenced by
the conditions there. Such regions may vary considerably, but the
dominant environment for Galactic star formation is one containing
thousands of young stars in a cluster of high density (typically
$10^4$~stars~pc$^{-3}$), and in which the spectrum of stellar masses
extends to bright O and B type stars -- from which they are usually known
as OB associations \cite{MilSca78,ClaBonHil00}. The presence of these
many other stars may have serious implications for the nascent
planetary system, as close stellar encounters can tidally disrupt the
protoplanetary disk, and UV radiation from the massive stars can
destroy it through photoevaporation.

The Orion Nebula Cluster (ONC), a nearby star-forming region only a few
million years old, is just such an environment. Containing over 4000
stars within a volume five parsecs across, it nestles at the edge of
the giant molecular cloud in Orion, about 470 pc away, and its
spectacular appearance has made it a familiar astronomical image.  At
its centre are the four bright Trapezium stars, of which the most
massive, \thoc, is by far the dominant source of UV
radiation in the cluster.

There is evidence to suggest that the majority of stars in the ONC have
circumstellar disks. Over 40 have now been observed directly with the
{\it Hubble Space Telescope}, either sillhouetted against the
background nebula or embedded inside a bright ionized envelope of
matter (a proplyd) \cite{MccOde96,OdeWon96,BalOdeMcc00}.  The existence
of many more can be inferred from an excess in the near infrared
continuum emission from stars in the cluster. Measurements by
Hillenbrand et al. \shortcite{Hil+98} found evidence for disks around
55--90 per cent of their sample of $\sim$ 1600 stars, and more
recently, Lada et al. \shortcite{Lad+**} have narrowed this fraction to 80--85
per cent in their observations.

In this paper we present the results of $N$-body simulations of the ONC
using Aarseth's {\sc nbody6} code \cite{Aar00}, and their implications
for disk destruction by photoevaporation and stellar encounters. Such
an approach is necessary to properly calculate the effects of
photoevaporation, because the mass loss rate from a disk due to an
incident UV flux is determined by its distance from the flux source.
The model we use for this, due to Johnstone, Hollenbach \& Bally
\shortcite{JohHolBal98}, St\"{o}rzer \& Hollenbach \shortcite{StoHol99}
and references in Hollenbach, Yorke \& Johnstone
\shortcite{HolYorJoh00}, is in agreement with measurements made by
Henney \& O'Dell \shortcite{HenOde99}, who found mass loss rates of
$\sim 4 \times 10^{-7}\,\msol\,\mbox{yr}^{-1}$ from proplyds close to
\thoc.  If these objects have spent the whole age of the ONC in this
environment, their initial masses must have been greater than
$0.8\,\msol$, and we might expect to see some disks today with a
non-negligible fraction of this mass present. However mm-wavelength
observations of disks in the ONC indicate \cite{Lad+96,Bal+98} disk
masses no greater than $0.02\,\msol$.\footnote{There are, however, many
uncertainties involved in these estimates. See Henney \& O'Dell
\shortcite{HenOde99}.} The paradox might be resolved if \thoc\ was born
only very recently -- at most $5 \times 10^{4}\,\mbox{yr}$ ago -- or if
the dynamics of the ONC were such that the proplyds seen close to
\thoc\ today have spent most of their lives elsewhere in the cluster,
as has been suggested by St\"{o}rzer \& Hollenbach
\shortcite{StoHol99}. We investigate this possibility in simulations in
which the cluster undergoes a collapse from cold initial conditions.

The relevance of cluster dynamics for the destruction of disks by
stellar encounters is even more readily apparent. A full dynamical
simulation is necessary to improve on analytic approximations in which
stars remain in the same density environment throughout the life of the
cluster (e.g. Clarke \& Pringle \shortcite{ClaPri91}).  Until recently
however, $N$-body codes suppressed strongly gravitationally focused
encounters by smoothing the gravitational field on small scales. In the
ONC, typical encounters are gravitationally focused at impact
parameters less than 160 AU, so a proper evaluation of systems on the
scale of interest for planet formation could not be made with such
codes.  In this work we use {\sc nbody6}, which represents the state of
the art in dynamical simulation on a commercial hardware platform, and
which incorporates two-body regularisation algorithms to handle close
encounters without smoothing.

\section[]{Composition and dynamics\\*of the ONC}

Although the ONC is relatively close, the large amounts of gas and dust
present and the bright emission from the Trapezium stars have until
recently hidden much of its population, and prevented us from
accurately measuring its kinematics. In the past decade, however, deep
and high resolution studies at optical wavelengths \cite{Pro+94} and
adaptive optical techniqes in the infrared \cite{MccSta94} have greatly
improved our understanding of the cluster's size and composition.

Hillenbrand \& Hartmann \shortcite{HilHar98} detected 3500 stars down
to mass $0.1\,\msol$ within $2.5\,\mbox{pc}$ of the cluster centre.  A
subsequent survey of the central $0.7 \times 0.7\,\mbox{pc}^2$
\cite{HilCar00} down to mass $0.02\,\msol$ found an additional 20~per
cent stars in that region. The overall mass spectrum was found to be
similar to that of the Galactic field. In this work we therefore
consider a population of 4000 stars, with a mean mass of around
$0.5\,\msol$ -- though the true population is probably higher still.

Jones \& Walker \shortcite{JonWal88} measured a (three-dimensional)
velocity dispersion $\sigma = 4.3 \pm 0.5\,\mbox{km}\,\mathrm{s}^{-1}$
for about 1000 stars distributed within $2\,\mbox{pc}$ of the centre.
Combined with a half-mass radius $R_{\mathrm{h}} = 1\,\mbox{pc}$, this
gives a crossing time $T_{\mathrm{c}} = 2R_{\mathrm{h}} / \sigma
\approx 0.5\,\mathrm{Myr}$.  The age of the cluster is difficult to
determine, but data from Hillenbrand \shortcite{Hil97} and Hillenbrand
\& Carpenter \shortcite{HilCar00} suggest that perhaps 85 per cent of
its stars are less than $2\,\mbox{Myr}$ old (with a mean age of $\sim
0.8\,\mbox{Myr}$). The cluster thus seems to be dynamically young -- a
few crossing times old.

Using these parameters, a rough calculation of the virial ratio
(kinetic energy / potential energy) of the cluster gives a value of
about 1.5. Some authors have concluded from this that either the ONC
must be unbound and expanding, or there must be a significant amount of
gas or stars -- perhaps low-mass binary companions -- present and not
seen \cite{JonWal88,Tia+96,HilHar98,Kro00}.  However, reasonable errors
in the observational parameters can easily account for an error of over
50 per cent in this calculation. The density profile of the cluster is
very much that of a relaxed system, to the extent that Hillenbrand \&
Hartmann \shortcite{HilHar98} were able to fit a King model --
characteristic of globular clusters many relaxation times old -- to
their data.

In addition, we note that even if the ONC is unbound today, it is
unlikely to have been so over much of its past evolution.  A highly
unbound cluster will always appear only about a crossing time old,
since it expands at some velocity $v \sim \sigma$, and its radius $R
\sim vt$, hence $T_{\mathrm{c}} \approx R / \sigma \sim t$. Also,
reasonable initial velocity distributions tend to evolve to a density
distribution which is much too flat when compared to that of the ONC
(Scally, in preparation).\footnote{Note that the asymptotic density
distribution of a highly unbound cluster is simply mapped from its
initial velocity distribution, since ultimately $r \approx vt$ for each
star. Thus an initial velocity distribution uniform in $v$ (say)
produces a density distribution uniform in $r$.}

\section{Proplyds and photoevaporation}

About 150 young stellar objects in the ONC have been observed with the
central star and disk surrounded by a bright extended ionisation front,
and a tail streaming in the direction away from a nearby massive star
-- usually \thoc.  These objects are known as proplyds (after
O'Dell, Wen \& Hu \shortcite{OdeWenHu93}), and in recent
observations \cite{BalOdeMcc00} about 80 per cent of the objects within
0.14 pc of the Trapezium stars are of this type. The fraction seems to
rise as one looks closer to the cluster centre \cite{OdeWon96}, though
this may be merely a selection effect, since their brightness drops in
proprtion to the square of their distance from \thoc, while the
background nebula fades less rapidly (O'Dell, personal communication).

A model for these objects has been developed by Johnstone, Hollenbach
\& Bally \shortcite{JohHolBal98} and St\"{o}rzer \& Hollenbach
\shortcite{StoHol99} (see Hollenbach, Yorke \& Johnstone
\shortcite{HolYorJoh00} for a review), in which their structure and
mass loss rate at a given distance from a massive star is determined by
the relative strengths of the star's far ultraviolet ($h\nu <
13.6\,\mbox{eV}$, hereafter FUV) and extreme ultraviolet ($h\nu >
13.6\,\mbox{eV}$, hereafter EUV) fluxes at that point.  Specifically,
they propose that there are two regimes in which photoevaporation
operates. In the inner one, FUV photons ominate the mass loss by
heating the circumstellar disk and causing a neutral flow out to an
ionisation front, where it meets the EUV field, as seen in the
proplyds. Within this regime the model predicts a mass loss rate
roughly independent of the distance $d$ from the massive star, since
the only criterion for it to apply is that the FUV flux is sufficiently
strong to heat the disk matter above its escape velocity.  In the EUV
regime, there is no neutral flow and the mass loss rate depends
directly on the EUV flux -- and hence on $d$. Since all proplyds
exhibit a stand-off distance between the disk and the ionisation front,
this model will produce them only in an FUV-dominated region.

Using the same physical parameters
as were assumed by St\"{o}rzer \& Hollenbach \shortcite{StoHol99},
Hollenbach, Yorke \& Johnstone \shortcite{HolYorJoh00} give the
following expressions for the mass lost by a disk in the FUV and EUV dominated regions:
\begin{eqnarray}
\dot{M}_{\mbox{\scriptsize FUV}} & \approx & 2 \times 10^{-9}r_{\mathrm{d}}\,\msol\,\mbox{yr}^{-1} \label{E:mdotFUV}\\
\dot{M}_{\mbox{\scriptsize EUV}} & \approx & 8 \times 10^{-12} r_{\mathrm{d}}^{3/2}\sqrt{\frac{\Phi_{\mathrm{i}}
}{d^2}}\,\msol\,\mbox{yr}^{-1}
\end{eqnarray}
where $r_{\mathrm{d}}$ is the disk radius in AU, $\Phi_{\mathrm{i}}$ is
the ionising (EUV) photon luminosity of the massive star in units of
$10^{49}\,\mathrm{s}^{-1}$, $d$ is its distance in pc, and we assume a
column density of $5 \times 10^{21}\mbox{cm}^{-2}$ from the ionisation
front to the disk inside a proplyd.

For \thoc, of type O6 \cite{Hil97}, $\Phi_{\mathrm{i}} = 2.6$, and the boundary
between the EUV and FUV dominated regions stands about 0.3 pc from the
star \cite{StoHol99}. The great majority of proplyds in the ONC have
indeed been observed within this distance, and it may be that those few
outside ($\sim$ 10 per cent in the survey of O'Dell \& Wong
\shortcite{OdeWon96}) involve a wind production mechanism other than
FUV heating. The other bright stars in the cluster are all less massive
and have significantly less UV output, with $\Phi_{\mathrm{i}} \approx
0.2$ for the second most massive (type O9), and $\Phi_{\mathrm{i}}
\approx 0.05$ for the third (type B0). Their FUV-dominated regions will
be correspondingly smaller ($\sim$ 0.1 pc and 0.05 pc respectively), as
will their photoevaporative effects.

The effect on planet formation is complicated by the fact that there is
a lower limit on the size to which disks can be reduced by the
photoevaporation process.  Ultimately this is determined by the mass of
the star, since for matter to escape from the disk at any point it must
be heated above the local escape velocity, corresponding to the
limit\footnote{The additional factor of $\sim 0.5$ is due to the
fact that material can actually escape from inside $GM_{*}/c_{\mathrm{s}}^{2}$
\cite{JohHolBal98}.}
\be
r_{\mbox{\scriptsize min}} \approx \frac{GM_{*}}{2c_{\mathrm{s}}^{2}}
\ee
where $M_{*}$ is the stellar mass and $c_{\mathrm{s}}$ is the sound
speed in the heated flow.  For a 0.3 \msol\ star (corresponding roughly
to the mode of the mass distribution in the ONC), the minimum radius in
the FUV-dominated region will be about 15 AU, while in the
EUV-dominated region evaporation can continue down to 1--2 AU -- the
difference being a consequence of the differing sound speeds in EUV and
FUV heated flows ($10\,\mbox{km}\,\mbox{s}^{-1}$ and
$3\,\mbox{km}\,\mbox{s}^{-1}$ resepctively) \cite{JohHolBal98}. Even in
an FUV-dominated region, once the disk becomes too small for FUV
radiation to drive a neutral flow, EUV-dominated mass loss will set in
for that system.

Within the minimum radius,
planets may well be able to form without hindrance, and for most
systems we may expect photoevaporation to have very little effect on
planets forming at about 5--10 AU from their central star (and none
whatsoever on planets in the inner 1 AU). But this central zone is
exactly the region we are most interested in for planet formation: the
planets in the Solar System are found there, as are many of the close
gas giants that have been discovered recently around nearby stars.

\section{Simulations}

We construct a dynamical model of the ONC, implemented using Aarseth's
{\sc nbody6} code \cite{Aar00}, consisting of 4000 stars, and starting
from a density distribution going as $r^{-2}$. Except where stated
otherwise (in the discussion of Figure \ref{F:coldfrac}), the results
shown refer to a cluster in virial equilibrium with a half-mass radius
of $\sim 1$ pc, and in all cases the initial conditions are chosen to
match the appearance of the ONC after an evolution time of 2--3 Myr (a
few crossing times). The mass function used is that of Kroupa, Tout \&
Gilmore \shortcite{KroTouGil93}, and the three most massive stars are
assigned the UV flux parameters specified in the previous section
($\Phi_{\mathrm{i}} =$ 2.6, 0.2 and 0.05). Of these, the most massive
is placed at the cluster centre, while the other two are given random
initial locations.  Several random realisations of this setup were
generated and run, but no statistically significant variations were
found, and the results presented here for one particular model are
characteristic of all those generated.

To model photoevaporation, we run two series of
simulations.  In the first we are interested
in the distribution of proplyds (in the FUV-dominated region) and for
every star in the cluster, so we keep track of three things during
the course of the simulation:
\begin{itemize}
\item the cumulative time it spends in the FUV-dominated region of any UV
source.
\item the time-integrated value of $\sqrt{\Phi_{\mathrm{i}} / d^2}$ for all three
UV sources whenever it is outside an FUV-dominated region.
\item the closest approach $r_{\mbox{\scriptsize close}}$ it makes to any
other star.
\end{itemize}
We then use the first two of these in (1) and (2) to evaluate the total
mass lost by each star due to photoevaporation during its life in the
cluster by assuming a fixed disk radius for all the systems throughout -
which we take to be 100 AU.

The assumption that the disk radii remain fixed throughout is a
simplification, since in reality it is likely that the disks will
decrease in size as they lose mass through evaporation. This would in
turn lead to a reduction in the mass loss rate, which scales with the
disk radius $r_{\mathrm{d}}$ in both the EUV and FUV regimes, as shown
in (1) and (2).  Johnstone, Hollenbach \& Bally \shortcite{JohHolBal98}
calculate the implied variation with time $t$ (at a fixed distance from
the ionising star) as $r_{\mathrm{d}} \propto t^{-1}$ (EUV) and
$r_{\mathrm{d}} \propto t^{-2}$ (FUV) for a disk whose surface density
goes as $r^{-3/2}$, and assuming that the disk is unable to replenish
the outer regions from which material evaporates.

Direct measurements of disk radii are difficult to make, and to a
certain extent depend on the wavelength one observes at. However,
estimates for various proplyds in the ONC \cite{JohHolBal98} vary from
20 to 80 AU, and are in agreement with the model's predictions taking
into account the size of the ionisation front stand-off and the
distance to \thoc\ in each case \cite{StoHol99}.  Given this, and the
possibility that the disks might have been much larger in the past if
they have already suffered a long period of photoevaporation, we take a
fixed value of 100 AU for these objects as a simplifying assumption
which will underestimate the total mass evaporated to date. It would of
course be possible to repeat our calculations while also keeping track
of the disk size for each system, but this would necessitate making an
assumption about their initial size -- as well as incorporating the
assumption that there is no viscous replenishment of the outer disk.
This would seem to be unwarranted in the light of our finding (later in
this section) that the mass loss assumption already implies
unacceptably large initial disk massses for the proplyds.

In the second series of simulations we are interested in mass loss from
the planet-forming region of the disk, i.e. the inner $\sim$ 10 AU. We
assume that there is no FUV-driven mass loss from this region -- which
is true for any star more massive than $\sim 0.2$ \msol\ (more than 60
per cent of cluster stars) -- so that for each star we only need to keep
track of the time-integrated value of $\sqrt{\Phi_{\mathrm{i}} / d^2}$
for all three UV sources. This is converted to a mass loss in (2) by
setting $r_{\mathrm{d}} = 10\,\mbox{AU}$. Once again the fixed disk
radius assumption is a simplification, which in this case will
overestimate the mass loss from the disk region we are interested in.
(We assume that the existence of disk material at radii greater than 10
AU does not increase the mass loss rate within that radius.)

\subsection{Photoevaporation}

\begin{figure}
\epsfxsize=8.0truecm\epsfbox{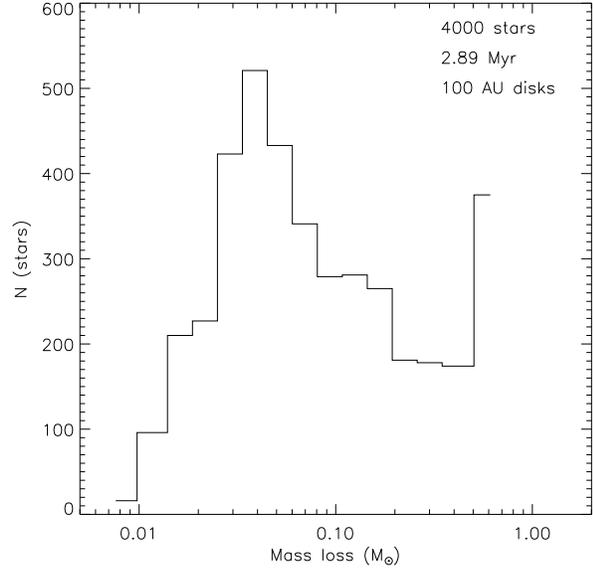}
\caption{Histogram of mass loss due to photoevaporation after 2.89 Myr
in the case where all disks have radius 100 AU throughout.}
\label{F:phothist100AU}
\end{figure}

Figure \ref{F:phothist100AU} shows a histogram of mass loss due to
photoevaporation after 2.89 Myr (roughly the ONC's current age) in the
case where the disks are 100 AU in radius.
We see that almost all systems have lost more than 0.01 \msol,
The rightmost bin contains systems which have spent
their whole time in {\thoc}'s FUV domimated region.

\begin{figure}
\epsfxsize=8.0truecm\epsfbox{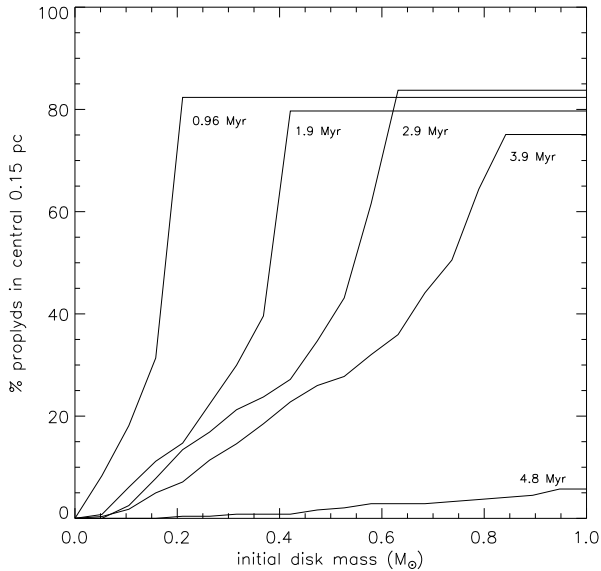}
\caption{Variation with initial disk mass of the percentage of
stars inside 0.15 pc (in projection) which are
proplyds (i.e. in an FUV-dominated region with some
circumstellar disk matter remaining) after 0.96, 1.9, 2.9,
3.9, and 4.8 Myr. All disks are assumed equal in mass initially and have
fixed radii of 100 AU. At 4.8 Myr the
percentage has dropped dramatically because we assume \thoc\ leaves the main
sequence at 4.6 Myr, after which only one or two less massive (but
longer lived) stars remain to ionize any disks.}
\label{F:propfrac}
\end{figure}

Figure \ref{F:propfrac} shows, at various times during the life of
\thoc, the percentage of stars
in the central projected 0.15 pc of the cluster which are proplyds, as a
function of the initial disk mass (assuming all disks are equal initially).
A proplyd in this context is a star in an FUV-dominated region with some
circumstellar disk matter remaining (and recall that the radius of
\thoc's FUV-dominated region is 0.3 pc). At the start of the simulation (or
immediately after the massive stars formed if that were different), all
stars in the central 0.3 pc (in 3D) would be proplyds, corresponding to
$\ga 80$ per cent of the stars within 0.15 pc in projection,
given the ONC's density distribution. But low mass disks
are very quickly destroyed, and to match the observed distribution of
proplyds at the ONC's present age requires a high initial disk mass
(0.4--$0.6\,\msol$ in Figure \ref{F:propfrac}) for these stars.
(Note that these initial disc masses would have to be even higher if we
were to include the effect of evaporation reducing the disk radii and
mass loss rates.)

\begin{figure}
\epsfxsize=8.0truecm\epsfbox{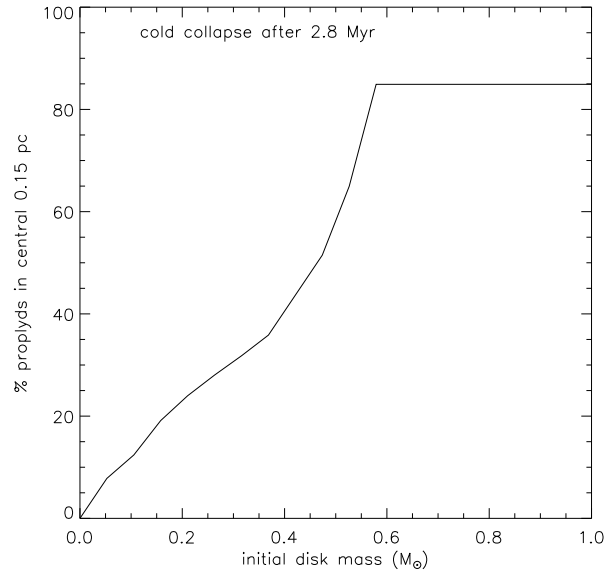}
\caption{As
Figure \ref{F:propfrac} but for an initally cold cluster after 2.8 Myr.}
\label{F:coldfrac}
\end{figure}

Figure \ref{F:coldfrac} is a similar plot showing data from a model
starting in an initially cold configuration, with potential energy
much greater than its kinetic energy. Such a system undergoes an
initial collapse, after which it virialises at about half its initial
size (subject to some contnuing small oscillation in radius). The
period of maximum density in the collapse is called the {\em crunch},
and models which have just passed through it can look similar
to the ONC. The initial density profile is important however, and
models starting from an $r^{-2}$ profile provide a better match to the
ONC at its present age than those starting from uniformity. In the
latter case, the core tends to be underdense, to the extent that only
about half the stars appearing in the central 0.15 pc in projection are
within 0.3 pc in 3D. Such a model can therefore never match the observations,
in which 80 per cent of stars in the centre are proplyds. The plot
shown here is for a cluster after 2.8 Myr with an intial
$r^{-2}$ density profile. It seems that
the dynamics have made little difference to how the fraction
of proplyds varies with initial disk mass, when compared to the case of
virial equilibrium. Stars which fall into \thoc's FUV-dominated region
during the collapse tend to stay there, and so for low-mass disks, the
population of depleted disks builds up even as new proplyds appear.

Note also that as time progresses, for any reasonable cluster dynamics,
the distribution of proplyds around \thoc\ should deplete from the
inside out, as these stars are more likely to have spent more time in
the FUV-dominated region. This assumes, of course, that there is no
other significant mechanism for creating proplyds, and that accretion
does not replenish the disks.

\begin{figure}
\epsfxsize=8.0truecm\epsfbox{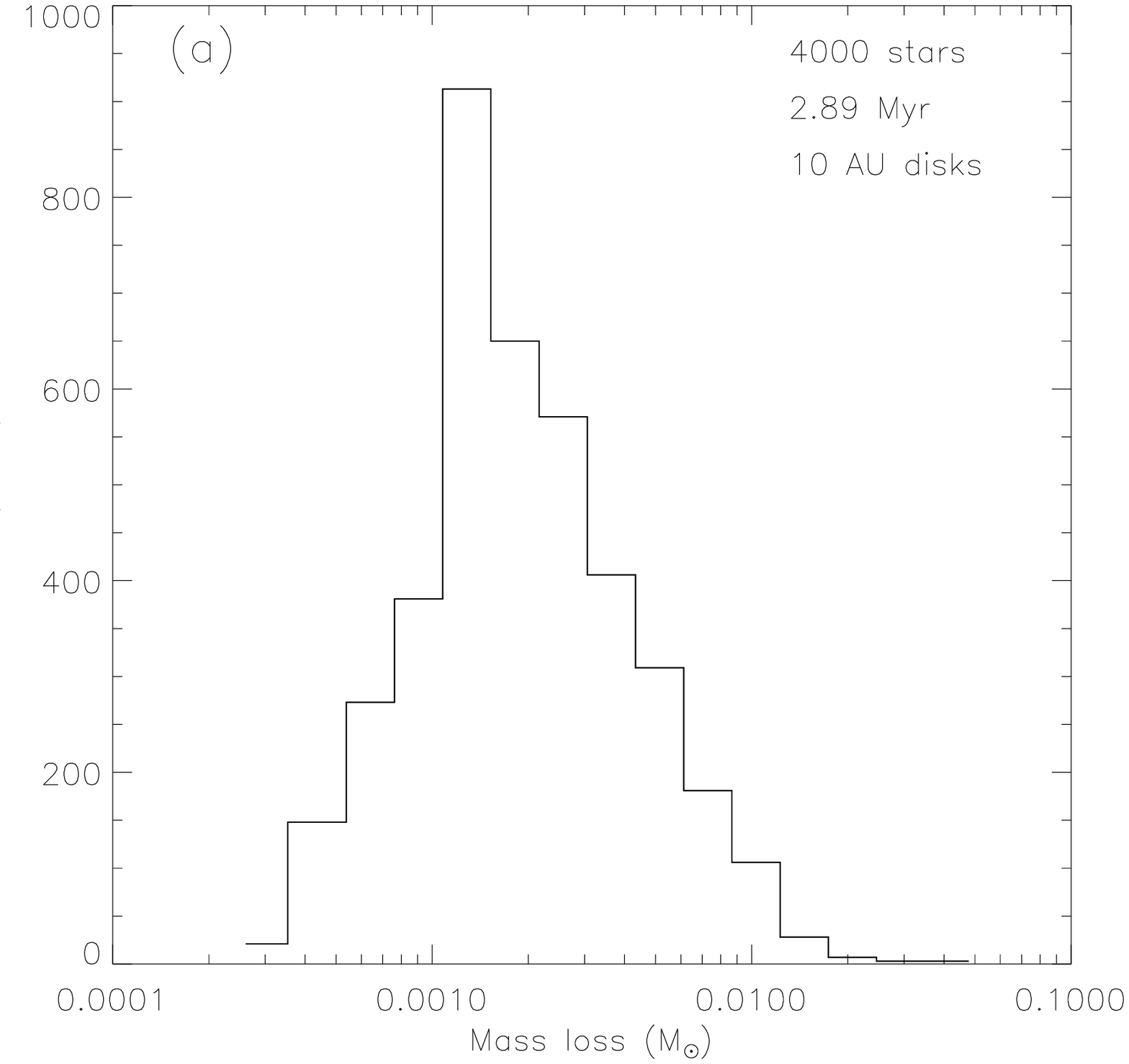}
\epsfxsize=8.0truecm\epsfbox{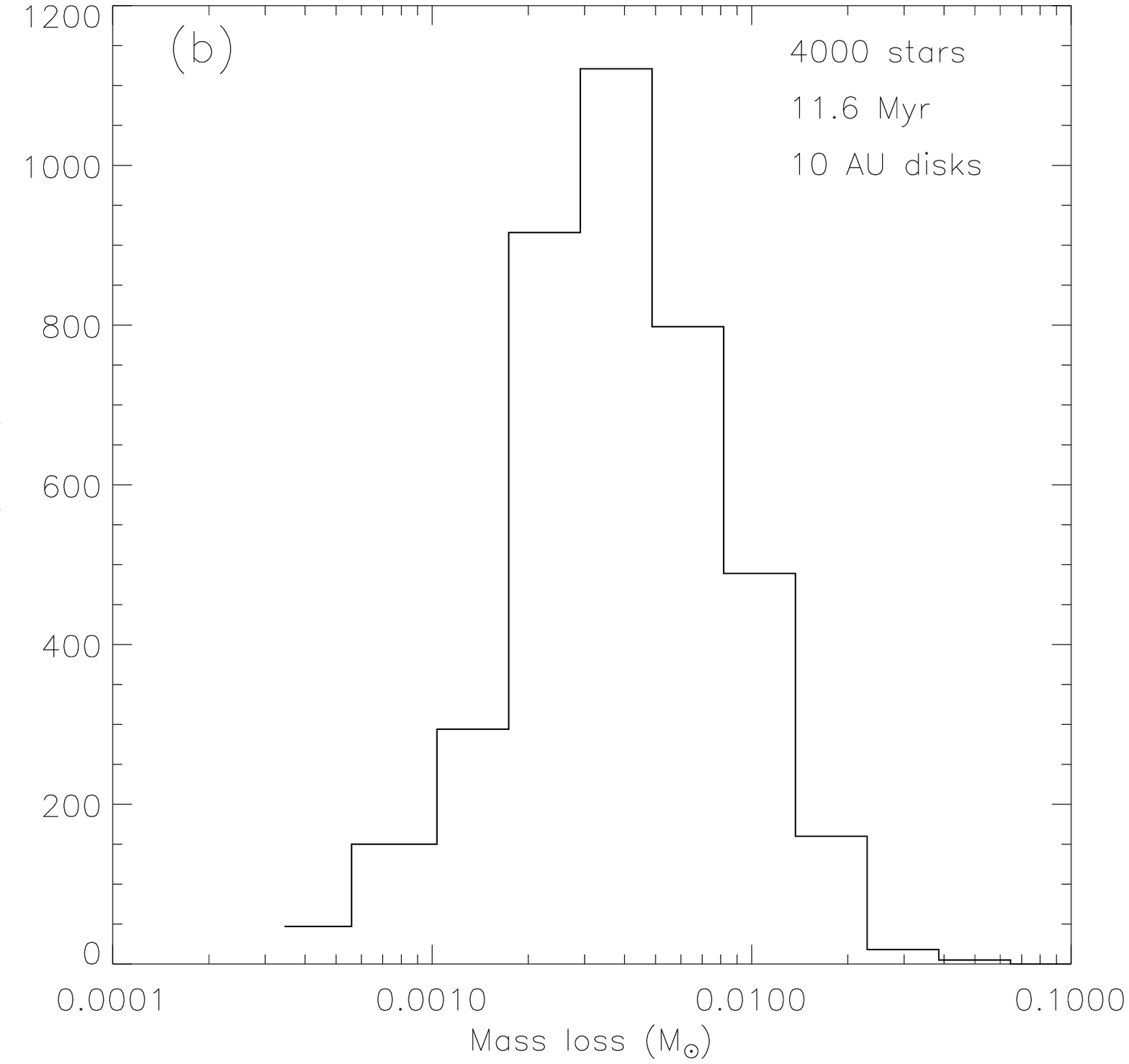}
\caption{Histograms of mass loss due to photoevaporation after 2.89 Myr
(a) and 12.5 Myr (b) in the case where all disks have radius 10 AU
throughout.}
\label{F:phothists10AU}
\end{figure}

Figure \ref{F:phothists10AU} shows histograms of mass loss due to
photoevaporation from 10 AU disks after 2.89 Myr (a) and 12.5 Myr (b),
by which time the massive stars will have left the main sequence and
photoevaporation will have finished. At this point, 95 per cent of
systems have lost less than the mass of the minimum Solar nebula
($\sim$ 0.013 \msol\ \cite{HayNakNak85}) (and recall that this is an
overestimate of the mass lost from the 1--10 AU region of their disks).
This is consisitent with the finding of Hillenbrand et
al.\ \shortcite{Hil+98} that in the centre of the ONC there is no
decrease in the incidence of disks revealed by near infrared excess.

\subsection{Encounters}

\begin{figure}
\epsfxsize=8.0truecm\epsfbox{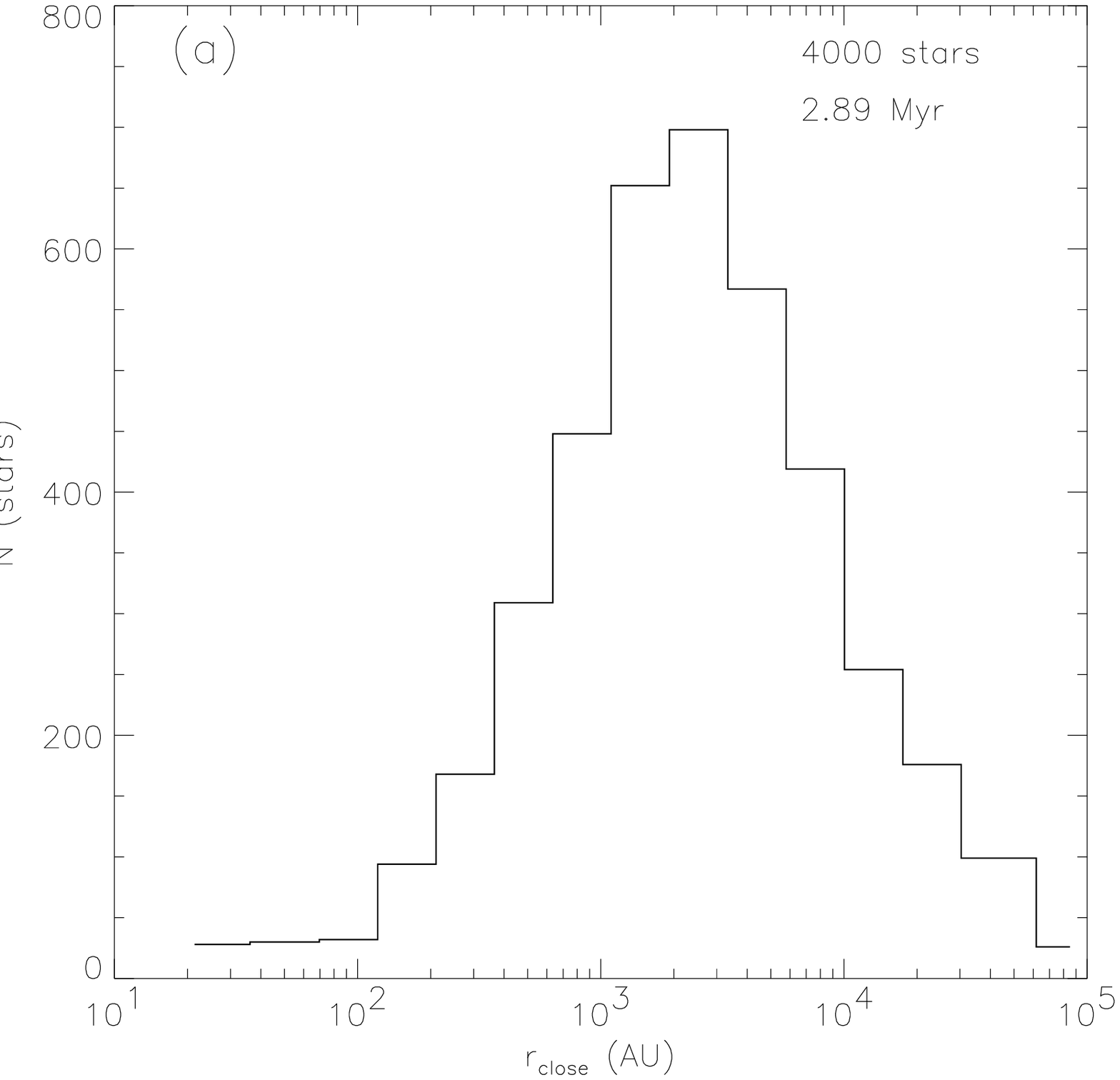}
\epsfxsize=8.0truecm\epsfbox{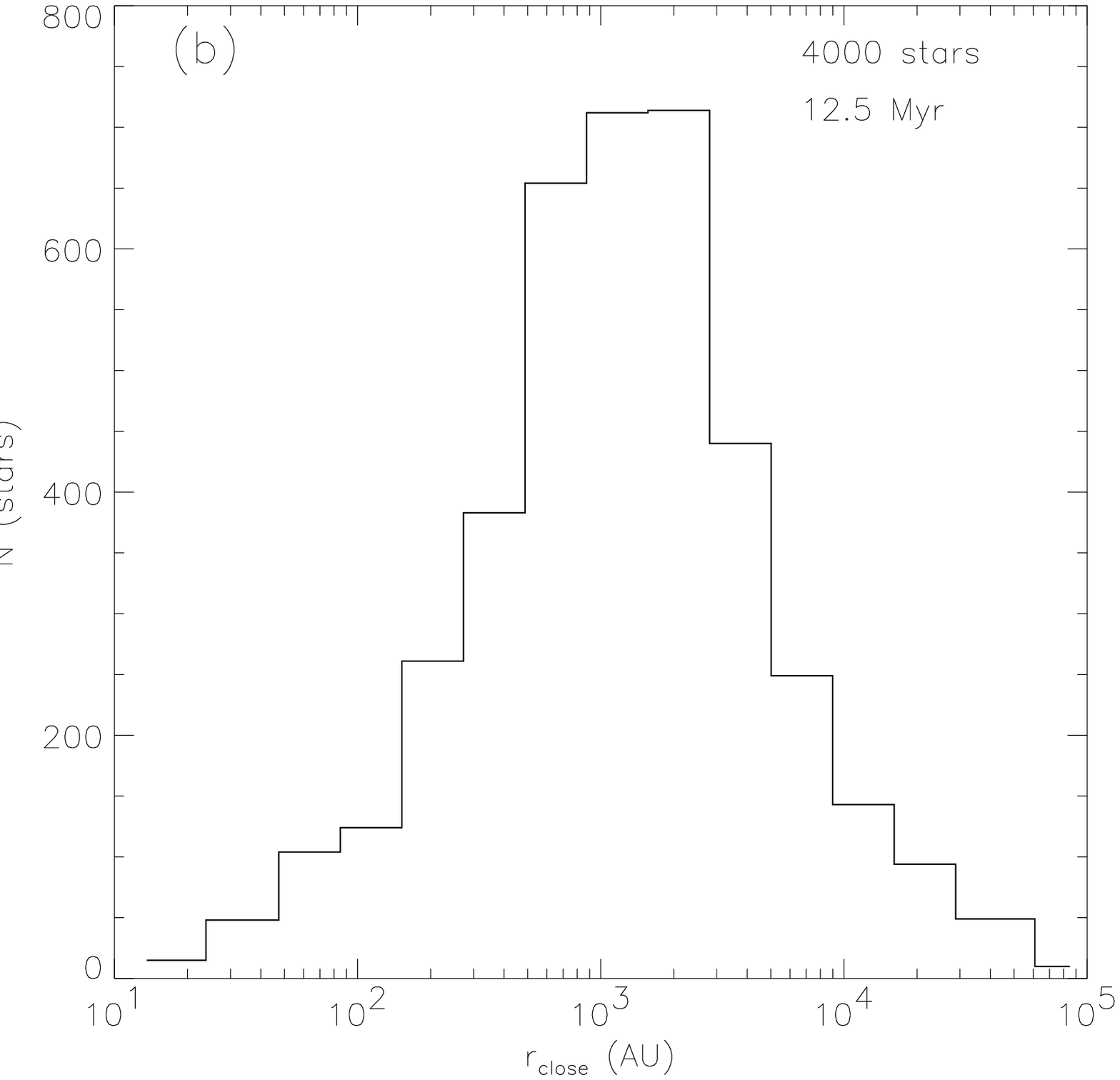}
\caption{Histograms of $r_{\mbox{\scriptsize close}}$ for
all the stars in the cluster after 2.89 Myr (a) and 12.5 Myr (b).}
\label{F:enchists}
\end{figure}

Figure \ref{F:enchists} shows histograms of
$r_{\mbox{\scriptsize close}}$ for all the stars in the cluster
at the ONC's present age and at 12.5 Myr. In this particular run we find that only about
3 per cent of stars have had an encounter 100 AU or closer at the present age,
rising to 6 per cent after 12.5 Myr. Typical values for these fractions,
averaged over a number of different runs, are $\sim$ 4 per cent and 8 per cent
respectively.

\begin{figure}
\epsfxsize=8.0truecm\epsfbox{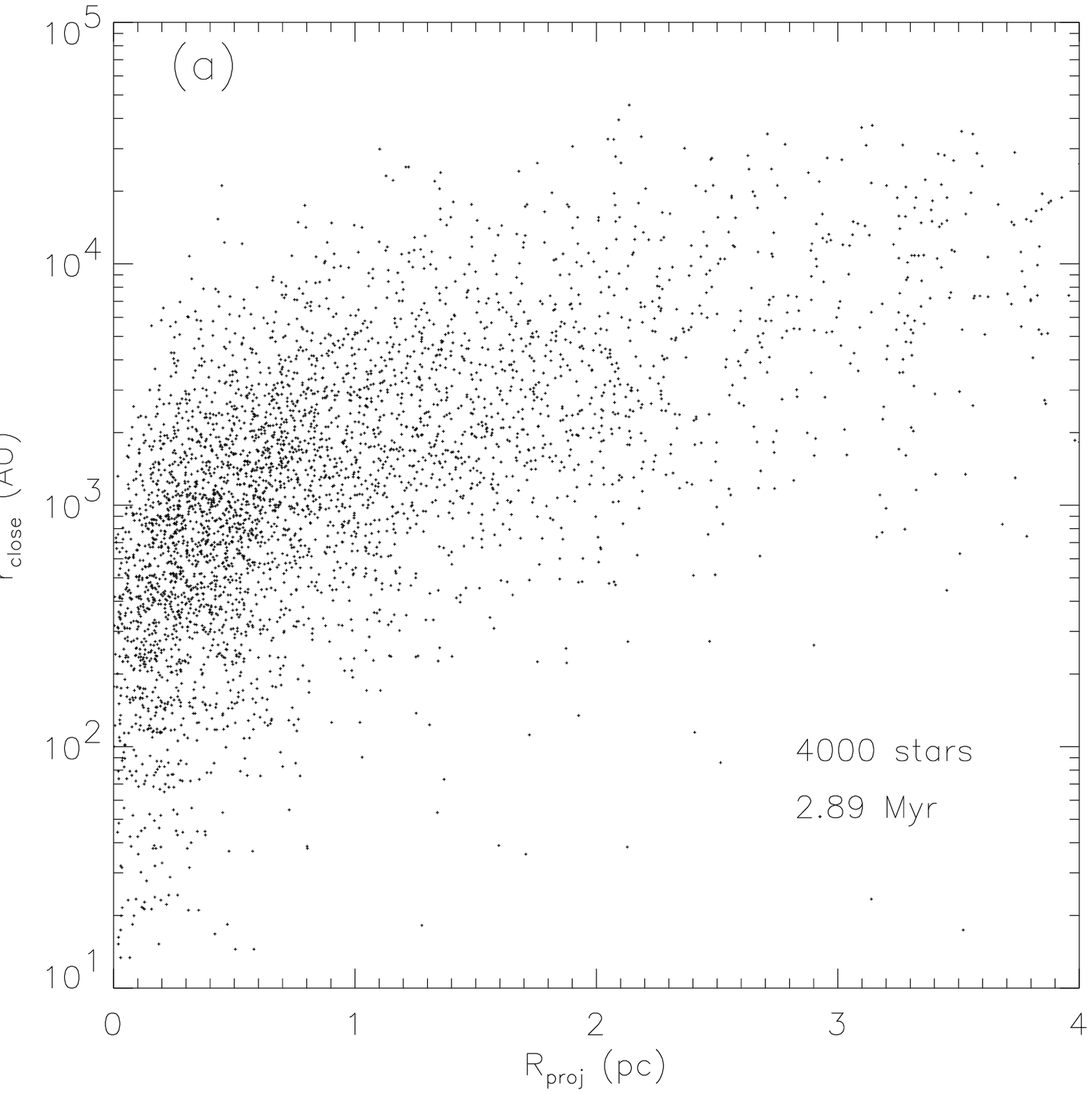}
\epsfxsize=8.0truecm\epsfbox{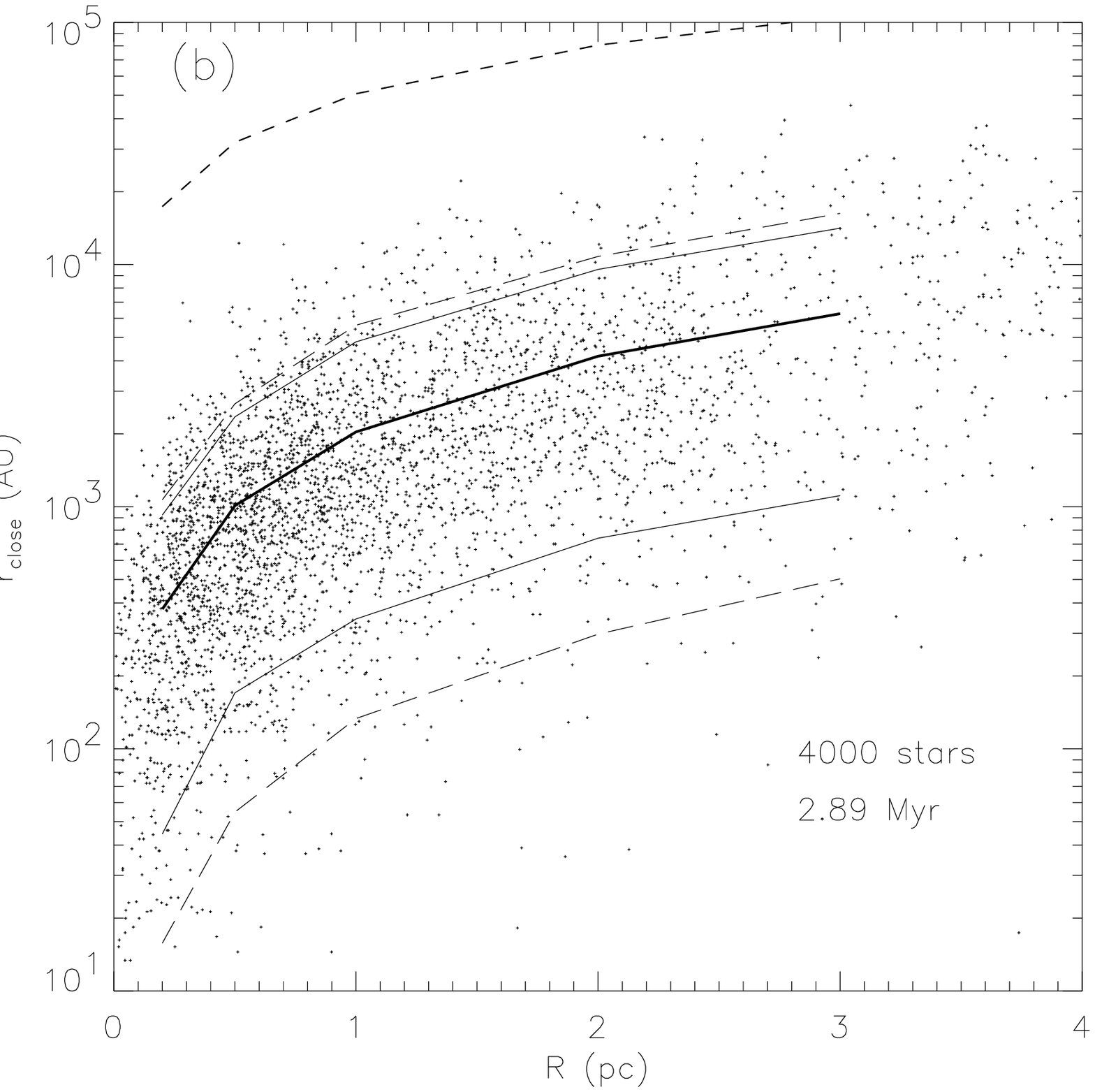}
\caption{(a) Closest encounter distance against projected
radial position after 2.89 Myr.
(b) Closest encounter distance against 3D radial position after
2.89 Myr. The thick line is the median of a distribution of closest encounters
produced by a Monte Carlo simulation (see Appendix) and the thin lines and
long-dashed lines are $2\sigma$ and $3\sigma$ limits respectively on
this distribution. The short-dashed line is the mean stellar
separation at $R$.}
\label{F:encplots}
\end{figure}

Figure \ref{F:encplots} shows all 4000 stars in the cluster on a plot
of $r_{\mbox{\scriptsize close}}$ against radial position, both in
projection and in three dimensions, at the ONC's present age. On the
plot in projection, we see no trend with $R_{\mbox{\scriptsize proj}}$
outside 1 pc from the cluster centre, with almost all stars having a
value of $r_{\mbox{\scriptsize close}}$ between $10^3$ and $10^4$ AU,
even though the stellar density drops by more than an order of
magnitude over the range plotted.

It is interesting to compare this with what one would expect if each
star had spent its whole time at the radius observed today -- i.e. if
no orbit in the cluster had any significant radial component.  To make
this comparison, we perform a Monte Carlo data simulation, the details
of which are presented in the Appendix.  In outline, the simulation
considers stars in radial bins, each bin having a local density which
determines the rate of encounters within any given distance. Encounters
are generated and assigned to the stars in the bin at random, and we
take the closest encounter for each star after the required time (2.89
Myr in this case, to compare with the dynamical simulation). This gives
a distribution in $r_{\mbox{\scriptsize close}}$ for each radial bin;
Figure \ref{F:encplots} plots its median and its $2\sigma$ and
$3\sigma$ limits against the data from {\sc nbody6}.

The Monte Carlo simulation matches the dynamical data remarkably well,
with the median, in particular, providing a good fit to the `centre' of
the distribution of stars. Its outer limits in $r_{\mbox{\scriptsize
close}}$ are narrower, and this may reflect the influence of orbital
mixing in the dynamical simulation, but the effect is minor, and we
conclude that simple analytical estimates give a good measure of the
encounter distribution at 2--3 Myr.

We find, however, that in evolving from $10^5$ to $10^7$ years, the
closest encounter distribution hardly changes. This is in marked
contrast to the analytic estimates, which imply that as a result of the
increase in available time the distribution should move to smaller
radii. For example, Bonnell et al.\ \shortcite{Bon+**} find that after
$10^7$ years, all of the systems in the core of a cluster like the ONC
(where the stellar density is $10^4\,\mbox{pc}^{-3}$) have had an
encounter within 100 AU, whereas in our models less than one third of
the systems in the core have had such a close encounter. The difference
between the two is mainly due to the fact that our model clusters
become significantly less dense over this timescale. This evolution is
not driven by mass loss or by super-virial initial conditions (though
these would have a similar effect), but is a consequence of the fact
that our initial conditions do not represent a steady-state solution of
the collisionless Boltzmann equation. The model clusters are generated
with a sharp discontinuity at the cluster edge, where the density drops
to zero.  In principle, the ONC could be the centre of a much more
extended distribution of stars comprising a steady state distribution,
as in the King model fit of Hillenbrand \& Hartmann
\shortcite{HilHar98},\footnote{Though such a cluster, to properly match
the relevant King model, would need to extend to about 20 pc from
\thoc\ and contain about 30,000 stars, which is almost certainly not
consistent with observations of the Orion region.} and in such a case
the density evolution would not occur and the core encounter
distribution predicted by Bonnell et al.\ \shortcite{Bon+**} would be
appropriate. However, we point out that for initial conditions that are
not so carefully constructed, we expect the core density to fall and
for most of the closest encounters in the ONC to have occurred by its
present age.\footnote{Another dynamical factor not present in the Monte
Carlo simulation is that for stars in the cluster core, many close
encounters may occur within small-$N$ groupings, rather than being
distributed over the entire core population.}

\section{Discussion}

\subsection{Proplyds}

It is clear from the mass loss rate in (\ref{E:mdotFUV}) that
photoevaporation can remove a large amount of disk material from
systems in the core of the ONC over the $\sim 5$ Myr lifetime of
\thoc.  Our simulations, which keep track of the stars' varying
distances from \thoc\ as they orbit in the cluster, allow us to
quantify the fraction of systems that never spend any significant time
near \thoc, and remain relatively unscathed by photoevaporation.  They
therefore shed some light on the much discussed problem of why proplyds
are common in the core of the ONC, when the short inferred survival
times of disks in this region would imply that most stars have already
have lost their disks and hence should not manifest proplyd activity.

One solution, proposed by St\"{o}rzer and Hollenbach
\shortcite{StoHol99}, is that the proplyds are merely `visiting' the
core region on radial orbits. In this picture, stars light up as
proplyds inside the cluster core, but since they have spent most of
their lives at much larger radial distances, their disks can have
survived for much longer.  We find no evidence for such a population of
stars on radial orbits in any dynamically plausible model for the ONC.
In models that are in virial equlibrium, the velocity dispersion is
initially isotropic, and remains approximately so, and the stars
currently in the central region have spent most of their lives there.
We find that in this case the observed high proplyd fraction is
consistent only with initial disk masses in excess of 0.4
\msol;\footnote{Which would of course constitute an unstable disk for
stars of Solar mass or less (see e.g.\ Laughlin \& Bodenheimer
\shortcite{LauBod94}, Toomre \shortcite{Too64}).} for somewhat smaller
initial disk masses, the proplyd distribution develops a ring shaped
structure in projection, as only disks towards the edge of the
FUV-dominated region (which spend some time outside it) can still
exist. The lack of any observational evidence for such a central
depletion in the proplyd distribution suggests that proplyds are not
close to the point of exhaustion.

Perhaps more remarkably, we find that the `radial orbit' solution of
St\"{o}rzer and Hollenbach also fails to be realised in the case where
the cluster undergoes cold collapse. In such models (where the kinetic
energy of the cluster is initially low, as would be the case if the
stars had fragmented out of a hydrostatically supported medium) the
stars begin by falling inwards on radial orbits. After about a
free-fall time, they achieve a configuration of maximum compactness
(the {\em crunch}) and rebound into a state of approximate virial
equilibrium. At an age of about 2 Myr, the ONC would have evolved
somewhat past the crunch and its orbits would be largely isotropic. In
consequence, the predicted proplyd fraction is remarkably similar to
the virialised case (Figure 3 compared with Figure 2).

Since we have shown that orbital dynamics cannot solve the proplyd
frequency problem, we are forced to invoke the other two solutions
considered by previous authors \cite{Bal+98,HenOde99}. The first
possibility is that the measured disk masses are vast underestimates.
In order to solve the proplyd frequency problem, the initial disk mass
required (0.4 \msol) would imply that 90 per cent of the stars only
lose a small fraction  of their disk mass to photoevaporation over the
lifetime of \thoc.  The second possibility (which would be compatible
with the low measured disk masses) is that \thoc\ has formed only
recently (implying incidentally that the ONC has been caught at a
special moment, and that older clusters containing OB stars would not
be expected to exhibit proplyd activity).

\subsection{Planets}

One might suppose that the high mass loss rates implied by (1) should
lead to a suppression of planet formation in all populous cluster
environments containing O stars, unless planets form at the same time
as the disk itself.  Certainly this is the case for planets forming
outside the inner 10 AU of the disk, since as Figure
\ref{F:phothist100AU} shows, even after 3 Myr most systems with 100 AU
disks will have lost an amount of disk material equal to several times
the minimum Solar nebula.  Planets would be prevented from forming on
such wide orbits in about half the systems in the ONC, unless, as
discussed above, either the initial disk masses were very high
(e.g.\ 0.4 \msol) or they formed quickly (in less than $\sim 2$ Myr)
and \thoc\ has only very recently appeared.

However, it is the inner 10 AU where we expect most planets to be found
(by analogy with our own system) and as explained in Section 2, only
EUV-dominated mass loss can affect this region of the disk. Figure
\ref{F:phothists10AU} shows that such mass loss is low -- relatively few
disks spend much time very close to an O star.  The results indicate
therefore that planet formation in the ONC would be largely unaffected
by photoevaporation.

Disruption due to stellar encounters is probably also unimportant. In a
star-disk encounter, matter can be stripped from the disk down to about
one third of the encounter separation \cite{ClaPri93}, and during the
lifetime of the cluster only a small minority of stars in our
simulations will have had encounters close enough to affect the
planet-forming region of their disks. This is essentially in agreement
with the results of Bonnell et al. (in press), who find that only in
the dense core of the ONC is there likely to be any noticeable
disruption to young planetary systems.

Our conclusion, therefore, is that in the absence of any disk
disruption mechanism other than those considered in this paper, we
would predict planets in orbits within 10 AU, and perhaps planetary
systems like our own, to be common amongst stars forming in ONC-like
environments.  As Armitage \shortcite{Arm00} has pointed out, in much
richer stellar environments, where the number of O stars is greater,
photoevaporation may have a more significant impact.  The observations
of Gilliland et al.\ \shortcite{Gil+**}, who find a complete absence of
close (`hot') Jupiter-mass companions in the globular cluster 47
Tucanae (which contains more than $10^5$ stars), may be an example of
this.  More challenging, perhaps, for our understanding of star and
planet formation are the results of the gravitational lensing survey of
Albrow et al. (in press), who find that less than a third of lensng
stars (typically of mass $\sim 0.3$ \msol) have Jupiter-mass companions
with orbits in the range 1.5--4 AU. If the majority of these lensing
stars were formed in an environment no richer than the ONC then neither
photoevaporation nor stellar encounters can explain the apparent
absence of planets.

\subsubsection{The Oort cloud}

We note finally that all stars in the ONC should have had an encounter
within a radius comparable to that of the Oort cloud in our own system
($\ga$ 20,000 AU), and that if the Oort cloud is primordial
(e.g. Cameron \shortcite{Cam73}), this could be used to rule out an origin
for the Solar system in an ONC-like environment.  On the other hand, it
is often supposed that the Oort cloud was formed from bodies scattered
out of the planet-forming zone by planetary and tidal perturbations
\cite{Oor50,Fer85}, arriving at large radii only after about $10^8$
years \cite{DunQuiTre87}.  If the ONC is unlikely to survive as a bound
cluster for that duration, its density would be considerably less by
then, and an Oort type cloud might be able to survive without
difficulty in the more dilute environment.  Given the uncertainties in
the origin of the Oort cloud, we conclude that its existence alone is
not sufficient to place any firm constraints on the birthplace of the
Sun.

\section{Acknowledgements}

We thank Matthew Bate, Sverre Aarseth, Ian Bonnell, Chris Tout, Bob
O'Dell, Lynne Hillenbrand and Mark McCaughrean for valuable help and
discussions. A. Scally is grateful for the support of a European Union
Marie Curie Fellowship.

\appendix

\section{Monte Carlo simulation of encounters}

In an environment with a given local stellar density
$n$ and one-dimensional velocity dispersion $\sigma$, the rate of
encounters (per star)
closer than a given distance $r_{\mathrm{e}}$ can be calculated as
\be
f = 4\sqrt{\upi}n\left(\sigma r_{\mathrm{e}}^{2} +
\frac{Gmr_{\mathrm{e}}}{\sigma}\right)
\ee
(e.g.  Binney \& Tremaine \shortcite{BinTre87}, p. 541)
where $m$ is the stellar mass, assumed the same for all stars. In a
cluster where the density $n(r) = kr^{-2}$ for some constant $k$
there are $4\upi k\,\delta r$ stars with radial positions in $[r, r +
\delta r]$, and after a time $T$ we expect them to have
\be
N_{\mathrm{e}}(r, \delta r, r_{\mathrm{e}}) = 16\upi^{3/2}k\,\delta r\,
n(r)T\left(\sigma r_{\mathrm{e}}^{2} +
\frac{Gmr_{\mathrm{e}}}{\sigma}\right)
\ee
encounters in total closer than $r_{\mathrm{e}}$.
Normalising this to $N_{\mathrm{e}}(r, \delta r, R)$ and inverting gives
a
generating function for stellar encounters closer than some maximum
distance $R$:
\be
r_{\mathrm{e}} = \sqrt{A^2 + BX} - A
\label{E:genfun}
\ee
where $X$ is a uniform random deviate in $[0, 1]$ and
\be
A = \frac{R + \frac{Gm}{\sigma^2}}{\frac{2R\sigma^2}{Gm} + 2}\;,\qquad
B = R^2 + \frac{GmR}{\sigma^2}
\ee
For $R$ we take the mean stellar separation at $r$:
\be
R = \left(\frac{48}{\upi n(r)}\right)^{1/3}
\ee
since all stars have encounters at this distance or greater at any
given instant.

Our procedure is then to generate $N_{\mathrm{e}}(r, \delta r, R)$
encounters using (\ref{E:genfun}) and assign them randomly to the
$4\upi k\,\delta r$ stars at radius $r$ in the cluster. Taking the
closest encounter for each star then gives a distribution of minimum
encounter distances at $r$, analagous to that obtained from the
dynamical simulation (with which we match the parameters $k$, $\sigma$,
$m$ and $T$).  Figure \ref{F:encplots} plots the median of the
distribution and its $2\sigma$ and $3\sigma$ limits.

\end{document}